\documentclass[aps,10pt]{revtex4}%
\usepackage{amsfonts}
\usepackage{amsmath}
\usepackage{amssymb}
\usepackage{graphicx}%
\providecommand{\U}[1]{\protect\rule{.1in}{.1in}}

\ifx\pdfoutput\relax\let\pdfoutput=\undefined\fi
\newcount\msipdfoutput
\ifx\pdfoutput\undefined\else
\ifcase\pdfoutput\else
\ifx\paperwidth\undefined\else
\ifdim\paperheight=0pt\relax\else\pdfpageheight\paperheight\fi
\ifdim\paperwidth=0pt\relax\else\pdfpagewidth\paperwidth\fi
\fi\fi\fi
\begin{document}
\title{The Spread of Viruses and Bugs in Self Organizing Networks }
\author{Swarbhanu Chatterjee}
\affiliation{Department of Physics, University of Rhode Island, Kingston, RI}
\affiliation{02881}

\begin{abstract}
We review recent progress made in analyzing the spread of viruses and bugs in
the internet. We describe how the use of a model that takes into account the
complex inhomogeneity of the internet and its self organizing characteristics
can lead to a better understanding of the persistence of some viruses compared
to others. We discuss how a better understanding can lead to a more targetted
antiviral and anti-bug solution.

\end{abstract}
\volumeyear{year}
\volumenumber{number}
\issuenumber{number}
\eid{identifier}
\date{June 4, 2011}
\startpage{1}
\endpage{ }
\maketitle

\section{Introduction}

The Internet is quickly becoming the backbone of our social and commercial
existence. The Internet makes possible what would have been impossible to
imagine only a few decades ago and its importance to the global economy can
hardly be overstated. There are two characteristics of the Internet that make
it uniquely important to physicists who study the laws of statistical
mechanics. The Internet is a self-organizing complex structure which seeks to
make itself more efficient within hardware constraints. This makes it very
similar to biological systems, more precisely living things, which self
organize with the specific aim of survival. Physicists have been trying to
explain the phenomenon of life for centuries using statistical mechanics but
have so far failed. It is here that the second attribute of the Internet makes
it a wonderful testing base for statistical mechanics. The Internet is
man-made and can be changed or experimented with extensively. We know
everything that is to be known about the small details of the Internet because
we have coded it ourselves. What we now need to investigate are the emergent
phenomena that are possible in the Internet. Understanding these emergent
phenomena may help us someday to understand the Holy Grail of science, the
phenomenon of life. Analyzing the spread of viruses and bugs may help us
someday know how cancer spreads in a living organism and why almost all animal
species on the planet are affected by it but not so much the plants.

The Internet can be considered as a scale-free network \cite{barabasi,amarel}.
A scale-free network is one in which the probability for a node to have $k$
connections to other nodes has a power law form,%

\begin{equation}
P\left(  k\right)  \sim ck^{-\gamma}, \label{scaling}%
\end{equation}
which therefore means that it has a long tail and nodes can always be found
that are heavily linked to other nodes. This power law form for $P\left(
k\right)  $ results from the fact that the Internet is a self-organizing
network which adapts in order to become more efficient. The mechanisms of
preferential attachment and fitness results in a situation when some nodes
that are well placed in the network or are simply more efficient begin to have
more links and become "super hubs". As a consequence of this tendency for the
Internet to have this inhomogeneous and not-completely-random structure, the
probability, $P\left(  k\right)  $, which by definition describes an average
effect, does not fall quickly as $k$ increases. Instead it has a power law
form. Such inhomogeneous networks in which there is a significant fraction of
nodes that tend to have high linkage are called scale-free networks. Li et.
al. defined a metric between $0$ and $1$,%

\begin{equation}
S\left(  g\right)  =\frac{%
{\textstyle\sum\limits_{\left(  i,j\right)  \in\epsilon}}
d_{i}d_{j}}{Max\left(
{\textstyle\sum\limits_{\left(  i,j\right)  \in\epsilon}}
d_{i}d_{j}\right)  } \label{metric}%
\end{equation}
which is maximized when the high-degree nodes are connected to other
high-degree nodes. Networks with low $S\left(  g\right)  $ are "scale-rich"
while those with high $S\left(  g\right)  $ are scale-free \cite{li}.

The Internet, while it connects people and businesses together, makes all
entities vulnerable to viral attacks or super bugs. Dealing with computer
viruses and bugs have become an ubiquitous part of urban life. In this paper,
we discuss the spread of viruses and bugs in the Internet
\cite{bug,epidemic,network,virusSpread}. Several governments, organizations
and experts are trying to make the Internet safer and more reliable
\cite{powergrid}. However, there has been a tendency to focus more on the
smaller scale of individual links, servers and computers. The antiviral
methods in use today focus on individual computers. We show in this paper that
such indiscriminate methods to stop the spread of viruses and bugs may not be
as effective as targetted methods can be view the Internet as a large
self-organizing network that adapts its complexity in a way as to always make
itself more efficient.

In order to analyze the spreading of a virus or a bug in a computer network,
it is useful to consider the following model. Each computer is a node and
every connection between two computers is a link. Therefore, a computer
network can be thought of as a network of nodes connected by links. The nodes
have only two states: healthy and infected. This is the standard
epidemiological model that has been used by Pastor-Satorras and Vespignani
\cite{epidemic} to describe the spread of a computer virus. At each time step,
a healthy node can be infected with a probability, $\nu,$ while an infected
node can be cured (because of antivirus software) with a probability, $\delta
$. The ratio, $\lambda=\nu/\delta$, defines a spreading rate.The authors have
assumed a \textbf{static model} in the sense that the parameters, $\nu$ and
$\delta\,$\ do not change in time. This is not true in the real world where,
as the authors have pointed out, antivirus software solutions are released in
a matter of days or weeks and individuals with once infected but now cured
computers become more wary of future attacks. Also individual browsing habits
change over time and therefore so does the network traffic. As individuals and
antivirus developers become aware of malicious sites, they block those sites
or do not visit them and therefore, the channels through which the viruses
spread are always evolving with time. We will come back to this point later
when we will discuss what happens when the network adapts or changes as is the
characteristic of self-organizing networks. In models with random graphs and
local connectivity, there is a threshold value, $\lambda_{c}$, below which for
$\lambda<\lambda_{c},$ the viruses decay exponentially fast. However, for
$\lambda>\lambda_{c}$, the virus spreads and becomes persistent. It has been
observed in real viral attacks of many different types, that viruses infact
saturate to a small but steady fraction of the total number of computers
connected to the Internet. Had random graphs been the real motif for the
Internet, such a result would have been surprising because there is no reason
to believe that all those viruses would have their spreading rate jut
infinitesimally above the threshold value. However, as the authors,
Pastor-Satorras and Vespignani, have shown numerically and analytically, a
large scale-free network has $\lambda_{c}\simeq0.$ This leaves the spreading
rate of the epidemic to be almost equal to the spreading rate of the virus,
$\lambda$, as set by the inventor of the virus. In their paper,
Pastor-Satorras and Vespignani noted the fact that the Internet and the World
Wide Web are scale-free networks with $P\left(  k\right)  \sim ck^{-\gamma}$.
To investigate the viral epidemics in scale-free networks, the authors built a
scale-free network numerically. They found following results:

1.The threshold, $\lambda_{c}=0$ for sufficiently large networks. However, for
networks of a finite size, $\lambda_{c}\neq0.$ They suggested that
fluctuations in the connectivity, ($\left\langle k^{2}\right\rangle =\infty$)
leads to infinite connectivity which makes it easier for the virus to spread
to the point of resulting in the threshold, $\lambda_{c}\rightarrow0 $.

2. They constructed a dynamical mean field reaction rate equation for the
relative density of the infected nodes, $n_{k}\left(  t\right)  \,,$ with $k$ links.%

\begin{equation}
\partial_{t}n_{k}\left(  t\right)  =-n_{k}\left(  t\right)  +\lambda k\left[
1-n_{k}\left(  t\right)  \right]  \Theta\left(  \lambda\right)
\label{meanField}%
\end{equation}
where, $\Theta\left(  \lambda\right)  $ is the probability for a link to point
to an infected node. In order to achieve a solvable mean field description,
they neglected the density correlations among infected nodes (i.e. $n_{k}^{2}$
terms). The steady state solution for the mean field distribution of the
infected nodes, $n\simeq\exp\left(  -1/m\lambda\right)  $, which meant that
$\lambda_{c}\simeq0.$ The fact that $\lambda_{c}\simeq0$ makes it easy to
explain the fact that viruses in the wild do not completely die off usually.
Instead, they stay alive in a small fraction of nodes. The fact that this low
prevalences of viruses in the wild can be explained by their model is an
important success of the model.

However, as noted before, the network considered by Pastor-Satorras and
Vespignani is static and does not change once it has been built
\cite{epidemic}. All the numerical simulations and analytical calculations
take place on a static network. This however is not true in reality and
therefore in this paper, we will provide an independent theoretical model
which will give an analytical explanation for how the self-organizing nature
of the Internet causes it to have characteristics similar to a true scale-free
network. The dynamics of a viral epidemic and the spreading of bugs in such a
self-organizing network is left to a following paper.

\section{The model}

In this section, we describe a very simple model that describes the essential
points in physics that determine the behavior of a large and evolving complex
network such as the Internet. The model does not claim to have all the
relevant mechanisms in it that can explain the behavior of the network.
Instead it has the barest minimum that can help to explain some of the not yet
completely understood characteristics of the Internet and provide us with an
intuitive handle with which to understand the Internet in a way that a more
complex analysis (that could be truly solved only numerically and not
analytically) would not be able to.

Before describing the model, we note that the assumption of Pastor-Satorras
and Vespignani regarding the static nature of the network is not really true
\cite{epidemic}. The hard-wired network consisting of personal computers,
servers, switches and routers connected via cables or wireless connections do
change in a time scale much longer than the time scale of a viral epidemic and
could be considered static. However, the "logical network" from the point of
view of a virus during its propagation is different than the above hard-wired
network and is this latter network adapts and evolves in a time scale that is
comparable or often less than the spreading time of the virus. To clarify
this, let us consider how a modern virus spreads vis-a-vis the old virus.
Earlier, viruses spread mostly through email attachments and this allowed the
viruses to be non-local because they could hop servers on the way but email
networks change over a few months or a year while the viruses itself can
spread over a matter of weeks. A\ modern virus however spreads mostly through
Internet advertisements and third party websites. These however adapt very
quickly, even on a daily basis, to the change in tastes of consumers. Besides,
anti-virus software solutions are released within days or weeks after the
first viral incident is discovered. Servers and computer networks that had
previously been susceptible to the virus suddenly is cured or becomes
impenetrable.This forces the viruses to adopt different website channels and
routes in order to propagate. The viruses are also more complex and use
different strategies to propagate. As a consequence, the "logical network"
through which a modern virus actually spreads is different than the actual
hard-wired network and evolves faster or atleast as fast as a modern virus
does. In the paragraphs that follow in this section, whenever we discuss a
network over which a virus spreads, we mean the "logical network". This
network is dynamic and not static as has been explained in this paragraph. In
fact, we will show later that the network is self-organizing, in the sense
that it evolves so that it becomes more efficient in terms of a certain
metric, which we will try to deduce.

Internet is mainly a channel through which information can be spread from
several sources to several sinks. What do we really mean by information? In
this paper, we define information in a most general way that is helpful enough
for our purposes. By information, we mean any meaningful data. A meaningful
piece of data is a sequence of characters (or more generally bits in computer
systems) that can be shown to be part of a language and which therefore when
processed can be "understood" by an automated system or a human in the sense
that it or he can do something with it now or in the future. The language may
be human or a computer language. We use the term language in the most general
sense that is possible to do so now. We note here that there is a considerable
amount of literature on the concept of a generalized language and generalized
semantics (please refer to Chomsky for early and defining work in the study of
languages). We denote information by a variable $\eta$ such that $0<\eta<1$.

In all problems in physics, a system can be described by a lagrangian and the
application of a variational principle in keeping with constraints always
leads to its equation of motion which describes the behavior of the system. In
what follows, we attempt to describe the flow of information along an evolving
large self-organizing computer network by a lagrangian so that its equations
of motions can tell us about the large scale behavior of the system. We can
visualize the Internet as a fractal structure of nodes and links. In general,
the fractal structure can have $D$ dimensions$.$ In reality, since the
Internet is inhomogeneous and its nature can be different in different
regions, the dimensions, $D$, can vary among regions. The location of each
node can be labelled by a coordinate, $x=\left\{  x_{1},x_{2}...x_{D}\right\}
$.

At this point, the simplest lagrangian for the information,$\eta\left(
x\right)  $,%

\begin{equation}%
\mathcal{L}%
_{\eta}=\frac{1}{2}\left(  \partial_{t}\eta\right)  ^{2}-f\left(  \eta\right)
\label{lagrangianInformation}%
\end{equation}
where the term, $\partial_{t}\eta,$ describes the rate of change of
information being held in a node (which typically is a server). The term,
$f\left(  \eta\right)  ,$is the inertia or potential energy for the
information carried by a node.%

\begin{equation}
f\left(  \eta\right)  =a_{0}+a_{1}\eta+a_{2}\eta^{2}+a_{3}\eta^{3}+...
\label{informationInertia}%
\end{equation}
where, $a_{i}$ are coefficients that may depend on the hardware resources. The
point is that in all nodes, the more information there is on the node, the
slower the node is in processing the information. This bottlenecking is an
ubiquitous feature in all traffic problems. The more saturated a node gets,
the more it chokes. Even when in the future, we will have servers with more
powerful processers and large memory banks, there is reason to believe that
our appetite for information of all kinds will increase manifold, and
bottle-necking will always be a problem. On the other hand, if we do have
resources that are more than enough for our demands, $f\left(  \eta\right)
\rightarrow0.$

So far we have considered only the dynamics of the information, $\eta\left(
x\right)  .$ But there is also the underlying computer infrastructure by which
we mean the "logical network" over which the viral epidemic spreads. Let the
variable, $d_{i},$ denote the number of linkages of this network. We define a
normalized function,%

\begin{equation}
\rho=\frac{d_{i}}{\max\left(  d_{i}\right)  } \label{normLinkage}%
\end{equation}
which is normalized, $0<\rho<1$.

Modifying the metric used by Li et.al. (\ref{metric}) which they maximized in
order to build scale-free networks, we get the following,%

\[%
\mathcal{L}%
_{\rho}=%
{\textstyle\sum\limits_{\left(  i,j\right)  \in\epsilon}}
b_{ij}\rho_{i}\rho_{j}%
\]
where, $\epsilon$ is the edge set and $b_{ij}$, are coefficients. For a
simplified treatment, we can consider the coarse-grained network, and in the
continuum limit,%

\[%
\mathcal{L}%
_{\rho}\left(  x\right)  =s\left(  x\right)  =b\left(  x\right)  \rho\left(
x\right)  ^{2}%
\]

where, $%
\mathcal{L}%
_{\rho}\left(  x\right)  $ is the lagrangian density for the network and
$b\left(  x\right)  $ is a constant in the simplest case. This lagrangian will
result in a clustering effect when the network is built. But there are no
terms in the lagrangian that can describe dynamic fluctuations, i.e. the
resultant network will be static and not self-organizing like the Internet is.

\bigskip The total lagrangian density that describes information flowing
through the network,%

\begin{equation}%
\mathcal{L}%
=%
\mathcal{L}%
_{\eta}+%
\mathcal{L}%
_{\rho}=\frac{1}{2}\left(  \partial_{t}\eta\right)  ^{2}-f\left(  \eta\right)
+b\left(  x\right)  \rho\left(  x\right)  ^{2} \label{totalLagrangian}%
\end{equation}

The lagrangian in Eq. \ref{totalLagrangian} describes the flow of information
through a large coarse-grained computer network such as the Internet. In its
present form, it can describe the cost that high amounts of information exacts
on performance of the servers that process the information. It can also
describe the clustering of the network.

\section{The self-organizing network}

We will now analyze the case of a self-organizing network. We consider the
network to be made of nodes and links that interconnect the nodes. In order to
describe the dynamics of the evolving network, we consider the nodes to be of
two kinds, super hubs (high performance servers) and weak nodes (low
performance computers). The state of a node is described by the variable,
$\xi=1$ for a super hub, $-1$ for a weak node. The hamiltonian of the system,
\begin{equation}
H=-J%
{\textstyle\sum\limits_{\left\langle ij\right\rangle }}
\xi_{i}\xi_{j}. \label{hubCoupling}%
\end{equation}
where, $J>0,.$determines the energy in the network. The free energy is given
by
\begin{equation}
F=\overline{E}-K\ S \label{freeEnergy}%
\end{equation}
where, $\overline{E}$ and $S$ are the mean energy and entropy of the system.
The parameter, $K$, plays the role that temperature plays in thermodynamic
systems. $K$ represents all the random stimuli that shake up the system much
as what heat does to a physical system. In the Internet scenario, there can be
several external stimuli that can try to randomize the system. These stimuli
include influx/outflux of money, government policies, politics and random
world events.

This system is very similar in several aspects to the Ising model with
ferromagnetic coupling but there is a very important difference. In the Ising
model, the lattice is static and Euclidean. Here, the lattice is fractal and
more over it is continually evolving. There is no inconsistency when
considering an evolving lattice. We are justified to apply some aspects of the
Ising model to the general case of a self-organizing network.

The coupling in the Hamiltonian, Eq. \ref{hubCoupling}, makes it advantageous
for nodes with large number of linkages connect to each other. Therefore, it
causes the formation of domains in each of which several nodes are next to
each other. These domains are separated by a sea of weak nodes. When $K$ is
small, then the mean energy of the system, $\overline{E}=-J%
{\textstyle\sum\limits_{\left\langle ij\right\rangle }}
\left\langle \xi_{i}\xi_{j}\right\rangle $, causes the formation of very large
domains of super hubs connected to each other. On the other hand, when $K$ is
large, the domains break into smaller domains.

Now we determine the entropy of the network. In general, the linkage in a
region of the network is $\rho\left(  x\right)  .$Consider a domain of
superhubs surrounded by a sea of weak nodes. Let the number of superhubs on
the boundary of the super hub domain which connect to the weak nodes outside
be $n.$Now consider a point on the boundary of the domain. The point can move
forward and backward along any link and create a different domain with a
larger or smaller size. Hence, since the coordination number of this lattice
is $\rho,$ an upper bound to the number of configurations of the boundary or
domain wall is $\rho^{n}$ \cite{goldenfeld}. This is a slight overestimation
since we have not taken into consideration that the boundary of a single
domain cannot intersect itself. However, $\rho$ is usually a large number much
greater than $1$ in our case and so the overestimation is not very
significant. Therefore, the entropy difference due to the presence of a domain
in that region of the network,
\begin{equation}
S\left(  \rho\left(  x\right)  ,n\right)  =\log\Omega\sim n\log\rho
\label{entropy}%
\end{equation}
where, $\Omega$ is the number of microstates. There can be more than one
domain of different sizes (and therefore, different number of super hubs at
the boundary, $n$) and each of them will contribute an entropy increase like
in the Eq. \ref{entropy}. The minimization of the total free energy,
$F_{tot}=\overline{E}-T%
{\textstyle\sum\limits_{i}}
S_{i}\left(  \rho\left(  x\right)  ,n_{i}\right)  $, in keeping with the
constraint for the total number of superhubs in the network, $N,$
\begin{equation}%
{\textstyle\sum\limits_{i}}
n_{i}=N, \label{numberSuperhubs}%
\end{equation}
results in a self consistent solution for the number of boundary nodes in the
$i-$th domain, $n_{i}=\overline{n}\left(  N\right)  $ and $\rho\left(
x\right)  $.

Therefore, the probability of a network to have $\rho$ linkages at any part of
it is,
\begin{equation}
P\left(  \rho\left(  x\right)  ,N\right)  \sim e^{-b\rho^{2}\left(  x\right)
}e^{S\left(  \rho\left(  x\right)  \right)  }\sim e^{-b\left(  x\right)
\rho^{2}\left(  x\right)  }e^{\overline{n}\left(  N\right)  \log\rho}%
=\rho^{\overline{n}\left(  N\right)  }\left(  x\right)  e^{-b\left(  x\right)
\rho^{2}\left(  x\right)  }, \label{prob}%
\end{equation}
where, $S\left(  \rho\right)  $ is the entropy in the network when there are
$\rho$ linkages in the network and the nodes of two possible kinds, super hubs
(high performance servers) and weak nodes (low performance servers). We should
note here that the issue of entropy in the network arises from the fact that
since the network is self organizing, linkages are constantly added, removed
or attached to other nodes. Therefore, in this evolving network, the nodes can
be connected together in different ways. Since the nodes can generally be
regarded to be of two kinds, high performance nodes and low performance nodes,
this results in there to be considerable amount of entropy in the system,
which has been discussed above.

We find that this distribution is not exactly the scale-free distribution, but
it has a longer tail than an exponentially falling distribution. The
distribution behaves in a similar way as that in a free scale network as far
as large numbers of linkages is concerned. But for small values, our
distribution starts as a power of $\rho$ instead of falling from a large value
as for a free scale distribution, $P_{free}\left(  \rho\right)  \sim
c\rho^{-\gamma},$which makes more physical sense. From Eq. \ref{prob},
$P\left(  \rho\left(  x\right)  \rightarrow1,N\right)  \sim\rho^{\overline
{n}\left(  N\right)  }$, while the free scale distribution, $P_{free}\left(
\rho\right)  \sim c\rho^{-\gamma}.$

\section{Summary}

We have formulated an independent theoretical model which provides an
analytical explanation for how the self-organizing nature of the Internet
causes it to have characteristics similar to a true scale-free network. In a
following paper, we will analyze the dynamics of a viral epidemic and the
spread of bugs in such a self-organizing network. A starting point would be to
include terms in the lagrangian corresponding to the virus or the bug, such as
$\left(  \partial_{t}\psi\right)  ^{2}$ and $\psi^{2}.$The second term leads
to a check in the population in the number of viruses or bugs. A reason for
the presence of this second term is that if there are too many viruses or
bugs, they will bring down the network or be detected quicker and antivirus
solutions will appear sooner to eradicate them. We note here that the main
difference between a virus and a bug is that a virus replicates and
regenerates while a bug is passive. This feature of viruses needs to be
reflected in the lagrangian model. In this paper, we have also not taken into
account the complexity of the nodes themselves (in the real world, the hubs
often grow in complexity as powerful processors are introduced
\cite{complexNode}).

Our models are simple to the point of being oversimplistic in some cases.
However, they are not meant to give exact accurate predictions regarding the
traffic in a network or the spread of a virus/bug. Instead, they are meant as
a tool that provides an easy to understand and intuitive handle to predict
certain kinds of large scale behavior of the networks and the viruses/bugs,
which numerical analyses rarely provide. Already the complexity of networks
today is at a point in which it is difficult to give accurate predictions
based on a numerical scheme. Also, there are not many ideas regarding how to
incorporate the self-organizing nature of the Internet into calculations of
virus epidemics and bug dispersion. As the Internet evolves into something
even bigger and more complex than what it is today, it may be useful to
simplify things and look at large scale behavior, which we have done in this paper.

\end{document}